\documentclass[12pt]{article}

\usepackage{graphicx}
\usepackage{times}
\usepackage{bm}
\usepackage{amsmath}
\usepackage{amssymb}
\usepackage{setspace}
\usepackage{color}
\usepackage{ulem}

\renewcommand{\Im}{\mathop{\rm Im}\nolimits}
\newcommand{\rmi}{{\rm i}}

\newenvironment{sciabstract}{%
\begin{quote} \bf}
{\end{quote}}

\title{Plasmon to exciton spin conversion in semiconductor-metal hybrid structures}

\author
{I.A.~Akimov,$^{1,2,\ast}$ A.N.~Poddubny,$^{2,\ast}$ J.~Vondran$^{1}$,  Yu.V. Vorobyov$^{1,3}$, \\ L.V.~Litvin$^{4}$, R.~Jede$^{4}$, \\ G.~Karczewski$^{5}$, S. Chusnutdinow$^{5}$, T.~Wojtowicz$^{6}$, and M.~Bayer$^{1,2}$
\\
\\
\normalsize{$^{1}$Experimentelle Physik 2, Technische Universit\"at Dortmund, 44221 Dortmund, Germany}\\
\normalsize{$^{2}$Ioffe Institute, Russian Academy of Sciences, 194021 St. Petersburg, Russia}\\
\normalsize{$^{3}$Ryazan State Radio Engineering University, 390005 Ryazan, Russia}\\
\normalsize{$^{4}$Raith GmbH, Konrad-Adenauer-Allee 8, 44263 Dortmund, Germany}\\
\normalsize{$^{5}$Institute of Physics, Polish Academy of Sciences, PL-02668 Warsaw, Poland}\\
\normalsize{$^{6}$International Research Centre MagTop, Institute of Physics,}\\ 
\normalsize{Polish Academy of Sciences, Warsaw, Poland}\\
\\
\normalsize{$^\ast$To whom correspondence should be addressed;}\\
\normalsize{E-mail: ilja.akimov@tu-dortmund.de, poddubny@coherent.ioffe.ru}
}

\date{}

\begin{document}


\baselineskip24pt


\maketitle


\begin{sciabstract}

\begin{singlespace}
Optical control of electronic spins is the basis for ultrafast spintronics: circularly polarized light in combination with spin-orbit coupling of the electronic states allows for spin manipulation in condensed matter. However, the conventional approach is limited to spin orientation along one particular orientation that is dictated by the direction of photon propagation. Plasmonics opens new capabilities, allowing one to tailor the light polarization at the nanoscale. Here, we demonstrate ultrafast optical excitation of electron spin on femtosecond time scales via plasmon to exciton spin conversion. By time-resolving the THz spin dynamics in a hybrid (Cd,Mn)Te quantum well structure covered with a metallic grating, we unambiguously determine the orientation of the photoexcited electron spins which is locked to the propagation direction of surface plasmon-polaritons. Using the spin of the incident photons as additional degree of freedom, one can orient the photoexcited electron spin at will in a two-dimensional plane.
\end{singlespace}

\end{sciabstract}

\section{Introduction}\label{Sec:Intro}

Optical initialization of electron spins is widely used to study the spin dynamics in a large variety of condensed matter systems~\cite{Dyakonov-book, Kirilyuk-Rev-2010}. The main advantage of optical control is the possibility of exceptionally fast manipulation with femtosecond light pulses which is attractive for applications in spintronic devices~\cite{Zutic-Rev-2008, Awschalom-book}. This approach relies on the transfer of photon spin to the electronic system due to absorption or scattering of the incident light. In particular, optical orientation of electron and hole spins by circularly polarized light is well established in semiconductors~\cite{Dyakonov-book}. Here, the photon spin projection $\pm\hbar$ onto the direction of its propagation dictates the spin orientation. 
However, it is much harder to establish optical orientation of photoexcited carriers in the plane perpendicular to the incident light, for which
only a few specific approaches are known. The first such approach is based on optical alignment of excitonic spin with linearly polarized light~\cite{OptAlignment}, requiring long-lasting exciton coherence which is observed only in a limited number of systems since the spin relaxation in the valence band is fast~\cite{Kosaka-PRL, Gershoni-2011, Langer-2014}. Linearly polarized light can also lead to optical orientation of electron spins in the presence of  exchange interaction with the nuclei~\cite{PRL-Dima} or when the electron spectrum is split due to reduced microscopic symmetry of  nanostructures~\cite{OO-LinPol-Tarasenko}.  However, none of these approaches can be used on timescales below 1~ps. Therefore, ultrafast optical orientation of photoexcited carriers with spin oriented perpendicular to the incident beam has remained an unresolved problem. 

Nanophotonic structures with dimensions comparable to the wavelength of light open unique possibilities for optical control of electron spins. First, the optical confinement within small volumes gives the opportunity to excite with nanoscale precision and high efficiency~\cite{Benson-2011}. Second, and even more important for optical spin manipulation, is the modification of polarization properties of the optical field confined within the photonic structure~\cite{Leuchs-rev-2015, Lodahl-et-al-2017,Bliokh-2015}. Here, propagating optical fields can possess large transverse spin locked to their propagation direction. Such spin fluxes were generated in  dielectric waveguides with properly positioned single quantum dots~\cite{Lodahl-2015, Makhonin-2016}. However, the inverse effect where the light spin flux would lead to optical initialization of carrier spins has not been demonstrated yet, which simply speaking would complement the establied field of magnetophotonics with the novel area of photomagnetics.

Plasmonic systems also support strong spin fluxes, enriching strongly the magneto-optical effects in hybrid metal-dielectric structure ~\cite{Temnov-2010, Belotelov-2011,  Akimov-2012, Maccaferri-JAP2020}. There, the optical response can occur very fast due to the extremely short lifetime of plasmons in the order of tens of fs ~\cite{Lineau-2005}. One of the most important excitations is the surface plasmon-polariton (SPP) propagating along the interface between metal and dielectric. Recently we demonstrated that spin conversion from excitons to SPPs leads to a significant enhancement of transverse magnetic routing of light emission (TMRLE), observed for a hybrid model system comprised of a magnetic semiconductor quantum well located in close proximity to a metallic grating that supports the SPPs~\cite{TMRLE-2018}. This effect required an in-plane magnetic field to excite spin polarized excitons, resulting in the highly directional emission via SPPs.
Here, we demonstrate the inversion of this effect, the spin conversion from SPPs to excitons on a femtosecond timescale.
The resulting optical spin orientation  in a direction perpendicular to the incident light is observed in the THz electron spin dynamics. Our results demonstrate that spin orientation can be optically achieved even without magnetic field. This is in stark contrast to the inverse magneto-optical Kerr effect which does not excite directly higher energy electronic states but rather deflects the already existing medium's magnetization from its equilibrium state~\cite{invTMOKE-Belotelov-2012, Chekhov-2018}.

\section{Plasmon to exciton spin conversion}\label{Sec:Idea}

\begin{figure}[t]
\centering
\includegraphics[width=.92\textwidth]{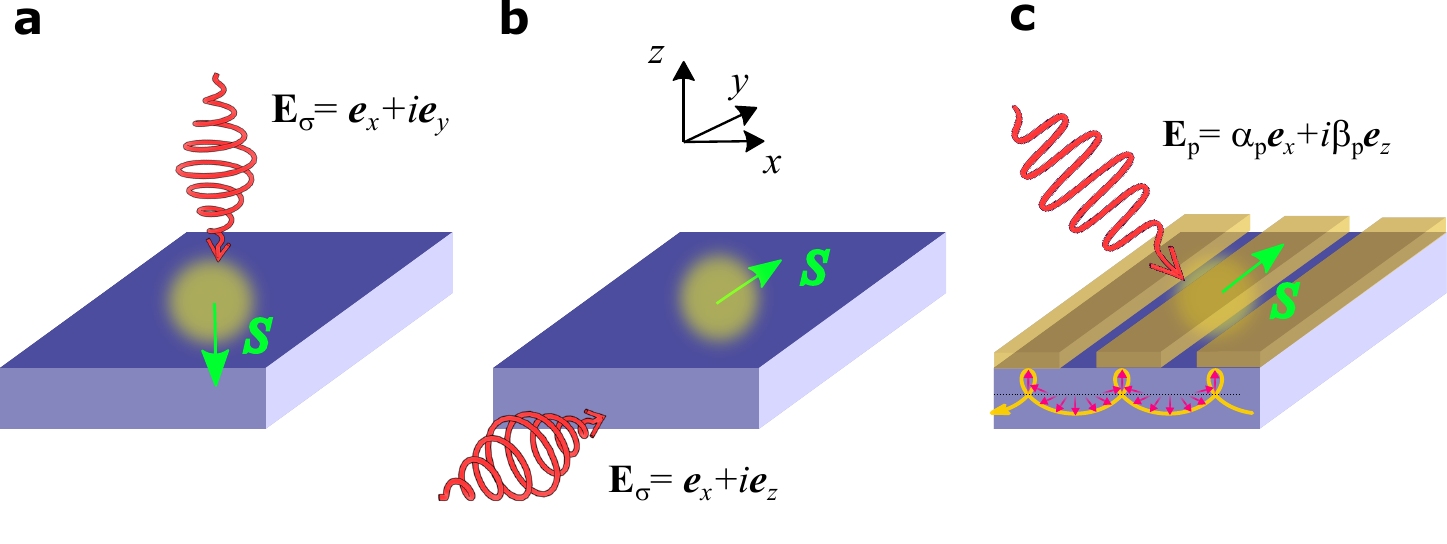}
\caption{\textbf{Optical orientation of electron spin} Conventional optical orientation of electron spin $\bm S$ along the $z$- (a) or $y$- (b) axes with circularly polarized light. (c) Optical orientation of electron spin along the $y$-axis via surface plasmon-polaritons (SPPs) with transverse spin which are launched by linearly $p$-polarized light.  The direction of the propagating SPP is indicated with yellow arrow while red arrows pointing towards the cycloid indicate the rotation of the SPP electric field vector $\bm{E_p$}.}
\label{Fig:scheme}
\end{figure}

In what follows, we present the basic approach for optical orientation of exciton (electron) spin in semiconductor nanostructures via spin conversion from SPP to exciton. Figure~\ref{Fig:scheme} shows three different geometries. For comparison, in panels (a) and (b) the conventional approach using a circularly polarized optical pulse is depicted where the electron spin $\bm S$ becomes oriented along the incidence direction of the light. The most widely exploited geometry uses normal incidence perpendicular to the sample plane (see Fig.~\ref{Fig:scheme}(a)). In this case the spin of the electron is directed along the chosen $z$-axis. It is also possible to orient the electron spin in the sample plane when the laser pulse is incident from the side (see Fig.~\ref{Fig:scheme}(b)), which often requires tight focusing and has limitations when the thickness of the active layer is smaller than the diffraction limit. 

Our approach is based on excitation of SPPs with energy $\hbar\omega_{\rm SPP}$ at the interface between a Au grating and a semiconductor. The periodic grating is required to launch SPPs which  propagate along the $x$-axis. Excitation is accomplished by linearly $p$-polarized light with photon energy $\hbar\omega$ under tilted incidence relative to the surface normal as shown in Fig.~\ref{Fig:scheme}(c). The grating period $a$ can be optimized to achieve most efficient coupling of the SPPs to the incident light, given by $k_x = k_{\rm SPP}\pm n\Lambda$ where $k_x$ is the wavevector component of the incident light along the $x$-axis, $k_{\rm SPP}$ is the SPP wavevector corresponding to the resonance condition  $\hbar\omega \approx \hbar\omega_{\rm SPP}$, $\Lambda$ is the reciprocal vector $2\pi/a$, and $n$ is an integer. SPPs propagating along the $x$ direction possess large transverse spin which is directed along the $y$-axis. The polarization of the SPP electric field at a certain position in the semiconductor close to the interface is given by $\alpha_p \mathbf{e_x} \pm \rmi\beta_p \mathbf{e_z}$ for the wavevector $\pm k_{\rm SPP}$, resulting in a circular polarization degree $\rho_p = \pm 2\alpha_p \beta_p $, where $\alpha_p,\beta_p$ are real parameters. The sign of polarization changes when the SPP propagates in opposite direction. Thus, the optical excitation of SPPs propagating along the $x$-axis in the semiconductor leads to optical orientation of electron spin along the $y$ axis at distances within the penetration depth of the evanescent SPP wave.

The well-defined selection rules for optical transitions in the semiconductor play an important role for quantitatively analyzing the spin control. We concentrate here on a (Cd,Mn)Te/(Cd,Mg)Te quantum well (QW) structure as model system, where the confinement in the $z$-direction significantly modifies the energy and spin level structure of the valence band. The choice of this QW as active layer where the electrons can be excited, is motivated by the possibility to locate it at a well-defined distance from the interface between metal and semiconductor.  However, our approach is generally not limited to some particular structure or material system. Conceptually, the same spin initialization will work for other semiconductors or condensed matter systems, where the spin-orbit coupling of electronic states (in our case in the valence band) allows one to transfer the SPP spin to photoexcited electrons.

\begin{figure}[t]
\centering
\includegraphics[width=.92\textwidth]{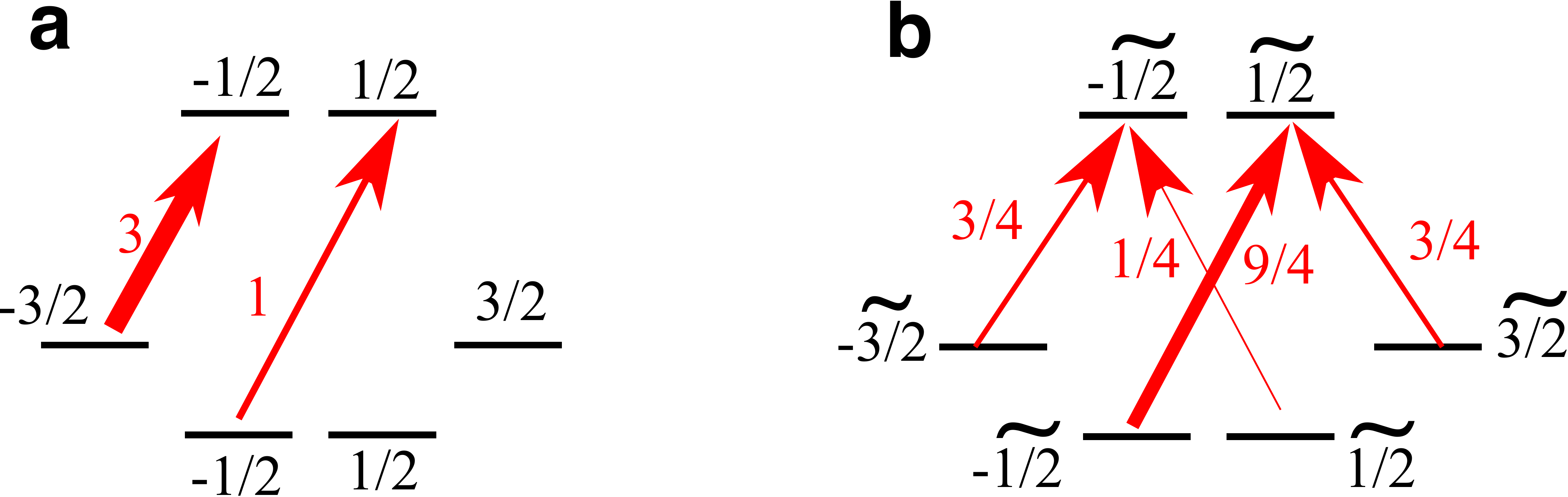}
\caption{\textbf{Selection rules for optical transitions}  
Schematic presentation of optical transitions and selection rules for excitation with circularly polarized light in a semiconductor quantum well. The valence band states with angular momentum projection of $\pm3/2$ (heavy holes) and $\pm1/2$ (light holes) are split due to confinement along the z-direction. The conduction band states have spin projections of $\pm1/2$. Arrows show excitation with  $\sigma^+$ circularly polarized light incident along the $z$ axis (a) and $y$ axis (b). For excitation along the $y$ axis, the rotated basis of eigenstates  
$|\pm \widetilde{1/2}\rangle$, $|\pm \widetilde{3/2}\rangle$ with the corresponding spin projections along the $y$-axis is used (for details see main text).
Numbers near arrows indicate the relative absorption probability, normalized in the same way for all transitions.}
\label{Fig:OpticalTransiitions}
\end{figure}

A schematic illustration of energy levels and selection rules for optical transitions is shown in Fig.~\ref{Fig:OpticalTransiitions}. We assume excitation with a spectrally broad femtosecond pulse which addresses both heavy- and light-hole states with angular momentum projections $\pm3/2$ and $\pm1/2$, respectively, which are split due to the QW confinement. The heavy hole spin is pinned to the confinement axis, which enables optical orientation of the heavy hole excitons only along the $z$-axis (see Fig.~\ref{Fig:scheme}(a)). The selection rules for excitation by circularly polarized light moving along the $z$ axis, perpendicular to the quantum well, enforce three times stronger heavy hole transitions compared to the light holes~\cite{OpticalOrientation}, see  Fig.~\ref{Fig:OpticalTransiitions}(a). As a result, the average spin polarization of the conduction band states is equal to $\langle S_z  \rangle= (-3+1)/(3+1)=-1/2$ after excitation. 
The selection rules for the transverse excitation geometry in Fig.~\ref{Fig:scheme}(b) with circularly polarized light propagating along the $y$ direction, are illustrated in Fig.~\ref{Fig:OpticalTransiitions}(b). We now use the basis of the conduction band states $|\pm\widetilde{1/2}\rangle$ with corresponding spin projections along the $y$ direction and also introduce the rotated states  $|\pm\widetilde{ 3/2}\rangle=(| 3/2\rangle\pm \rmi |- 3/2\rangle)/\sqrt{2}$ for the heavy holes  and $|\pm\widetilde{ 1/2}\rangle=(|1/2\rangle\pm \rmi | -1/2\rangle)/\sqrt{2}$ for the light holes. The dipole moment of the heavy-hole excitons lies in the quantum well plane ~\cite{IvchenkoPikus}, and thus they can be excited by $\bm e_{x}\pm i\bm e_z$-polarized photons with the same efficiency as in the normal geometry and do not contribute to optical spin orientation in the $y$ direction, see  Fig.~\ref{Fig:OpticalTransiitions}(b). 
However, spin polarization is still possible due to the light hole excitons, which have non-zero dipole matrix elements in all directions. Namely, the probability of optical excitation of the light hole exciton with electron spin  $S_y=+1/2$ by transverse circularly polarized light is nine times larger than that for electron spin  $S_y=-1/2$, see Fig.~\ref{Fig:scheme}(b). Summing up all transitions from light and heavy hole states, we obtain the transverse electron spin polarization degree $\langle S_y \rangle= (9-1)/(9+1+3+3)=1/2$, which is the same as $\langle S_z \rangle$ in the geometry of Fig.~\ref{Fig:scheme}(a). Thus, the selection rules enable omnidirectional electron spin initialization, both perpendicular to and in the quantum well plane, depending on the excitation geometry.  

The same analysis applies for the plasmon to exciton spin conversion, shown in Fig.~\ref{Fig:scheme}(c). The average electron spin polarization in the quantum well plane is given by $\langle S_y \rangle=\rho_c/2$, where $\rho_p = \pm 2\alpha_p \beta_p $ is the transverse spin polarization of the plasmon. Hence, excitation of the structure by $p$-polarized light in the $xz$ plane launches plasmons, initializing electron spin along the $y$ direction. This initialization protocol establishes different electron populations with opposite spin projections, and does not require coherences between electron and hole states. As such, it is immune to fast hole spin relaxation, contrary to the approach in Refs.~ \cite{Kosaka-PRL,Langer-2014}. Moreover, transverse spin can be initialized   even by unpolarized light for oblique incidence, because the sign of $S_y$ is locked to the plasmon propagation direction and does not directly depend on the incident light polarization.

\section{Pump-probe in magnetic field}\label{Sec:Result}

This concept of plasmon to exciton spin conversion was tested for a 10~nm thick diluted-magnetic-semiconductor Cd$_{0.974}$Mn$_{0.026}$Te QW structure which was previously used for demonstration of magnetic field-induced routing of the exciton emission~\cite{TMRLE-2018}. The QW layer was seperated from the surface by a 32~nm thick Cd$_{0.73}$Mg$_{0.27}$Te barrier. The surface of the structure was partly covered with rectangular field containg gold gratings (period 250~nm, slit width 50~nm, and Au thickness 45~nm) using electron-beam lithography and subsequent lift-off processing (see Methods for details). A part of the sample was left uncovered which allowed us to compare the electron spin dynamics between the bare semiconductor and hybrid plasmonic QW structures.

\begin{figure}[t]
\centering
\includegraphics[width=0.8\textwidth]{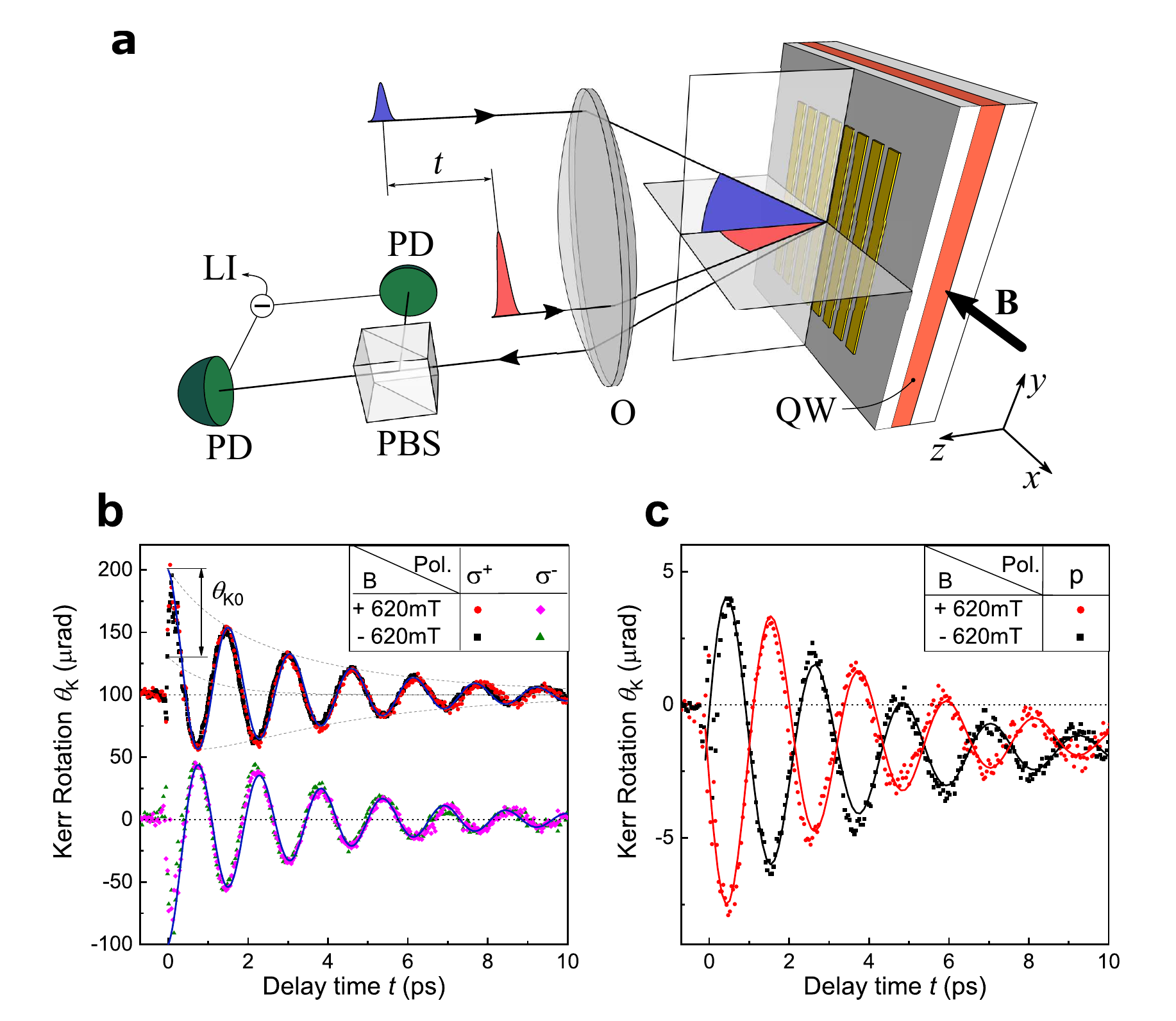}
\caption{\textbf{Experimental realization and spin transients}
(a) Scheme of pump-probe experiment (O is objective, PBS is polarization beam-splitter, PD is photodiode and LI is lock-in amplifier). The pump and probe beams are incident in horizontal and vertical planes, respectively. Pump-probe transients measured for bare QW (b) and hybrid plasmonic (c) structures using circularly polarized and $p$-polarized pump, respectively.}
\label{Fig:Transients}
\end{figure}

To determine the direction of optically oriented electron spins and to resolve their dynamics, we perform transient pump-probe Kerr rotation in a transverse magnetic field (see Fig.~\ref{Fig:Transients}(a) and Methods). The sample is kept at low temperatures $T=2$~K. The ultrashort 30 fs laser pump pulse centered at $\hbar\omega=1.64$~eV photon energy with a spectral width of about 0.1~eV excites both heavy- and light-hole QW excitons at energies of 1.66 and 1.68~eV, respectively. The pump beam is incident at an angle of $\Theta_{\rm Pump}\approx 10^\circ$ and its plane of incidence is perpendicular to the grating slits, as required for excitation of directed SPPs in the hybrid plasmonic structure. The linearly polarized probe pulse hits the sample after the delay time $t$ relative to the pump, at an angle of 7$^\circ$ in the plane parallel to the slits. Its reflection is direcetd into the polarization sensitive balanced photodetection setup. In our experiments, the rotation of the probe polarization plane by the angle $\theta_K$ due to the polar Kerr effect is dominant (see supplementary section I for evaluation of the detected probe signal for different polarization configurations). Therefore, the pump-induced signal is given by $\theta_K(t) \propto S_z$, where $S_z$ is the photoexcited electron spin polarization along the $z$ axis.

The pump-induced initial spin polarization of the electrons, $\bm{S_0}$, and its subsequent dynamics in the external magnetic field $\bm{B}$ depend strongly on the polarization of the pump pulse, type of structure (bare semiconductor or hybrid plasmonic) and magnetic field direction. In the bare QW structure only circularly polarized light leads to the excitation of a non-zero spin $S_z$. The sign of $S_z$ changes when the pump helicity is switched between $\sigma^+$ and $\sigma^-$. In the hybrid plasmonic structure excited with a $p$-polarized pump, we expect to induce an in-plane spin polarization $S_y$. For the magnetic field directed perpendicular to the slits, spin precession around $x$-axis with the Larmor frequency $\nu_L=g_e\mu_B B/h$ will takes place in both of the cases described above, where $g_e$ is the electron g-factor, $\mu_B$ - Bohr magneton and $h$ - Planck constant. The signal is given by damped oscillations,  $\theta_K \propto S_z(t) = S_0 \exp(-t/\tau_S) \cos (2\pi\nu_L t + \varphi)$, where the phase $\varphi$ is defined by the initial spin orientation $\mathbf{S_0}$ and $\tau_S$ is the electron spin relaxation time.

Figures~\ref{Fig:Transients}(b) and \ref{Fig:Transients}(c) show characteristic pump-probe transients measured for the bare QW under $\sigma^+$ and $\sigma^-$ pump excitation and the hybrid plasmonic structure for a $p$-polarized pump, respectively. The transients are shown for magnetic fields applied in opposite directions with the magnitude $|B| = 640$~mT. The spin dynamics in the bare QW is well described by cosine-like oscillations with $\varphi \approx 0$, in agreement with previous studies in diluted magnetic semiconductor QWs~\cite{Crooker-97}. The initial sign of the spin polarization $S_0$ depends on the helicity of the polarization and the dynamics is obviously independent from the direction of magnetic field. Moreover, we do not observe any measurable spin dynamics under excitation with linearly polarized light which indicates very fast hole spin relaxation (see supplementary section 3). In stark contrast to the bare semiconductor, for the plasmonic structure we observe that a linearly $p$-polarized pump leads to optical spin orientation in the QW plane. This is manifested by the oscillatory signal with $\varphi \approx \pm \pi/2$ for $B = \pm 640$~mT. The change of the phase from $+\pi/2$ to $-\pi/2$ when the magnetic field is reversed clearly demonstrates that the pump induced spin is excited along the $y$ axis. For an $s$-polarized pump the oscillatory signal becomes negligibly small, in agreement with our analysis, as because no SPPs are excited in this case (see supplementary section 2 for spin dynamics dependence on linear pump polarization). Note that the oscillations are superimposed on a step-like signal, which is not related to the photoexcited electron spin dynamics and can be most likely attributed to heating (for details see supplementary section 1). Thus, the comparison of pump-probe transients in Fig.~\ref{Fig:OpticalTransiitions}(b) and Fig.~\ref{Fig:OpticalTransiitions}(c) clearly shows the plasmon-to-exciton conversion in the hybrid plasmonic-semiconductor QW structure. 

\begin{figure}[t]
\centering
\includegraphics[width=0.6\textwidth]{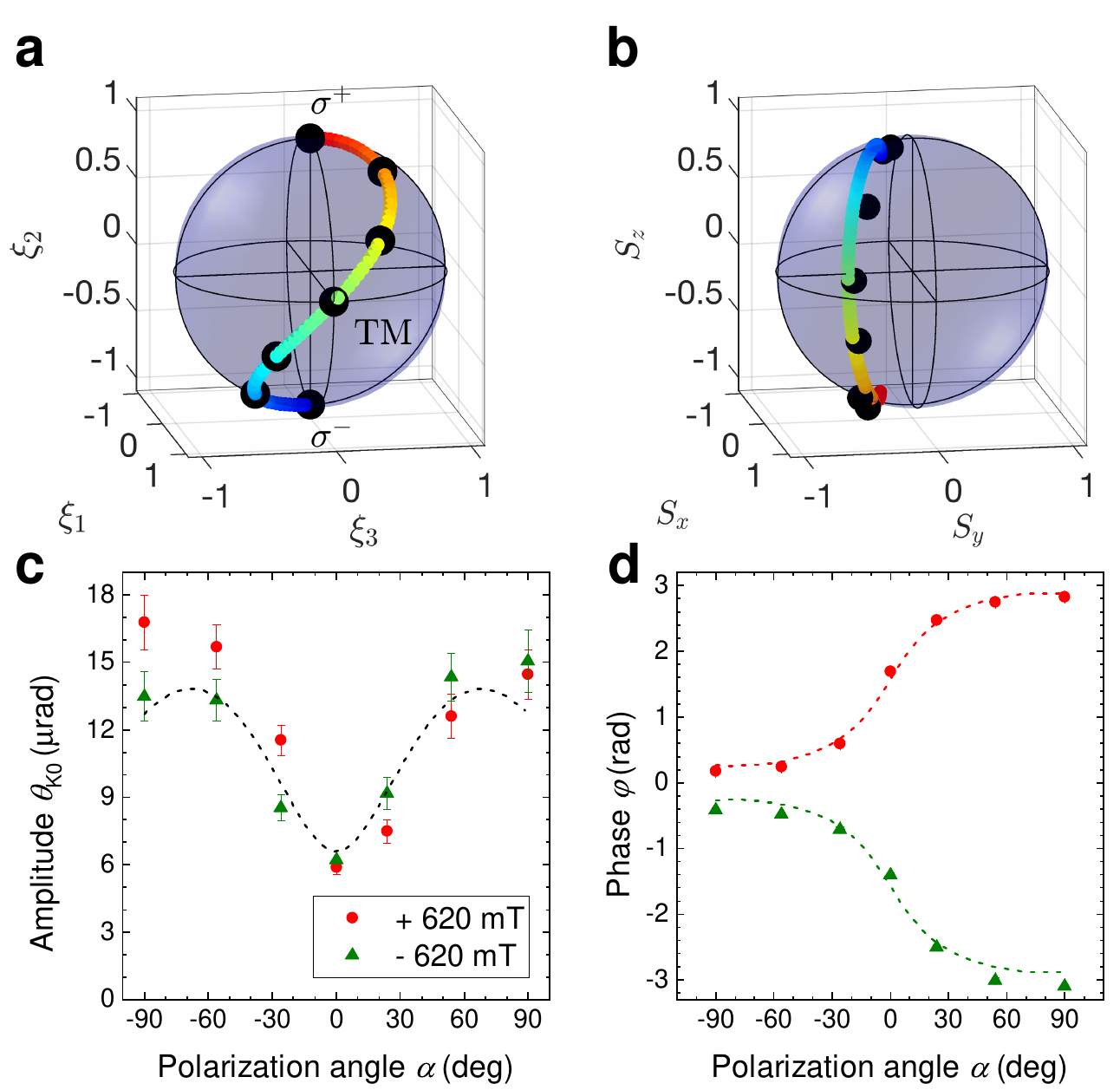}
\caption{\textbf{Optical orientation in hybrid structure for different polarization states of pump excitation}  
The polarization of pump excitation pulse and resulting spin orientation $\bm{S_0}$ are shown on Poincare (a) and Bloch (b) spheres; the $\xi_i$ are the Stokes parameters. Colored lines and dots correspond to theory and experiment, respectively. (c) and (d)  Initial amplitude $\theta_{K0}$ and phase $\varphi$ of Kerr rotation transients in hybrid plasmonic structure plotted as function of angle $\alpha$ defining the polarization state of the pump pulse as shown in (a). Data are presented for magnetic fields $B = \pm 620$~mT.}
\label{Fig:PolaDep}
\end{figure}

Special attention deserves the excitation with mixed polarization which simultaneously contains $p$- and $s$-linear polarized components with well-defined relative phase. In this case, both the photoexcited SPPs and the elliptically polarized plane wave transmitted through the grating contribute to optical orientation of the electron spin. Each mechanism itself initializes the spin, but along different axes and thus allowing for omnidirectional spin control in the $yz$-plane. 

These results are summarized in Fig.~\ref{Fig:PolaDep}. For experimental realization we used a $\lambda/4$ plate with the main axis set by an angle $\alpha/2$ with respect to the direction of the initially linearly polarized pump pulse. In that way, the polarization state of the incident pump could be varied from $\sigma^+$ ($\alpha=-90^\circ$) to $\sigma^-$ ($\alpha=+90^\circ$) with linearly polarized $p$-wave ($\alpha=0$) in the intermediate state as shown on the Poincare sphere in Fig.~\ref{Fig:PolaDep}(a) (for details see also Methods).
The corresponding initial electron spin at $t=0$ is shown on the Bloch sphere in Fig.~\ref{Fig:PolaDep}(b) and is evaluated theoretically from the Kerr rotation transients, i.e. the initial amplitude $\theta_{K0}$ and phase $\varphi$ which dependences on the polarization angle $\alpha$ are shown in Fig.~\ref{Fig:PolaDep}(c) and (d), respectively. 
In the calculation the electric field inside the structure was expanded as \cite{Whittaker1999}
\begin{gather}\label{eq:E}
\bm E(x,z)={\rm e}^{\rmi k_xx}[\bm{\mathcal{E}}^s_{\rm far}(z) E^{(0)}_s +\bm{\mathcal{E}}^p_{\rm far}(z)E^{(0)}_p]+\bm{\mathcal{E}}^s_{\rm near}(x,z)E^{(0)}_s+\bm{\mathcal{E}}^p_{\rm near}(x,z)E^{(0)}_p\:,
\end{gather}
where $k_x=\omega \sin\Theta_{\rm Pump}/c$ is the in-plane wave vector of the incident wave, $E^{(0)}_{s}$ and $E^{(0)}_{p}$ are the polarization components of the incident wave.
The far-field components $\bm{\mathcal{E}}^{s,p}_{\rm far}(z)$ depend on $x$ in the same way as the incident wave, $\propto {\rm e}^{\rmi k_xx}$.  The near-field components 
 $\bm{\mathcal{E}}^{s,p}_{\rm near}(z)$ are modulated by the grating and their coordinate dependence is described by the wave vectors $k_x\pm 2\pi/d,\quad k_x\pm 4\pi/d,\ldots$. The $s$-polarized field $\bm{\mathcal{E}}^s$ is polarized along $y$ while the $p$-polarized field $\bm{\mathcal{E}}^p$ has nonzero projections on both the $x$ and $z$ axes. In our numerical calculation the field in Eq.~\eqref{eq:E} was rigorously calculated using the Fourier modal method~\cite{Whittaker1999,Li1996}.
 The exciton spin $\bm S_0$ induced by the pump was determined from the electric field Eq.~\eqref{eq:E} as 
 \begin{equation}\label{eq:S}
 S_{0,z}=a_{hh}\Im \langle E_xE_y^*\rangle,\quad  S_{0,y}=a_{lh}\Im \langle E_xE_z^*\rangle,\quad S_{0,x}=a_{lh}\Im \langle E_yE_z^*\rangle\:,
 \end{equation}
 where the angular brackets denote the spatial averaging over the coordinate range $x=0\ldots d$ in the unit cell and the coefficients $a_{hh}$, $a_{lh}$ describe the relative efficiencies of  excitation via heavy- and light-hole transitions, respectively. 
In addition we take into account in Eq.~\eqref{eq:S} that the efficiency of light hole excitation is smaller in the QW due to the complex valence band structure including heavy-light-hole mixing for energetically lower lying states, $a_{lh}<a_{hh}$. In particular, it is well established that the electron spin polarization in QWs for excitation via the light-hole like states is always smaller as compared to the realitvely pure heavy-hole exciton ground state~\cite{Weisbuch-81,Oestreich-2005}. 
Thus, we  use the ratio of the heavy- to light-hole excitation efficiency $\eta=a_{hh}/a_{lh}$ as a fitting parameter. 

It follows from Fig.~\ref{Fig:PolaDep}(d) that excitation of the plasmonic structure with a circularly, $\sigma^+$ or $\sigma^-$, polarized pump gives $\varphi \approx 0$ and $\pi$, respectively. In this case the initial electron spin is oriented along the $z$ axis, i.e. the main contribution comes from the conventional mechanism of spin excitation as in the bare QW, $S_{0,z}\propto \Im E^{(0)}_s E^{(0)*}_p$. For the $p$-polarized pump, $\varphi \approx \pm \pi/2$, so that $\bm {S}_0$ is excited mainly along the $y$-axis by the SPP spin flux, $S_{0,y}\propto\Im \mathcal E^{p}_{{\rm near},x} \mathcal E^{p *}_{{\rm near},{z}}\ne 0$. The phase dependence on the angle $\alpha$ changes continuously from 0 to $\pi$, implying rotation of the electron spin direction from $+z$ to $+y$ and further to $-z$ and manifesting omnidirectional spin control. 

The dependence of the Kerr rotation angle $\theta_{K0}(\alpha)$, shown in Fig.\ref{Fig:PolaDep}(c), is proportional to the initial spin $S_0(\alpha)$ and allows us to evaluate the efficiency of spin orientation by each of the two mechanisms. The magnitude $\theta_{K0}$ for $\sigma^+$ pump excitation is about twice larger as compared with a $p$-polarized pump. The ratio depends on the competition between the far-field and near-field contributions in Eq.~\eqref{eq:E}. We also find that good agreement with the experimental data is obtained for the ratio of heavy and light hole excitation efficiencies $\eta = 10$ (see solid curves in Fig.~\ref{Fig:PolaDep}(c) and (d)). The ratio between the SPP and far field wave mechanisms can be varied by adjusting the design of the plasmonic part of the structure, which can be used to tune the details of the initial spin dependence $\bm{ S_0}(\alpha)$.

\begin{figure}[t]
\centering
\includegraphics[width=.3\textwidth]{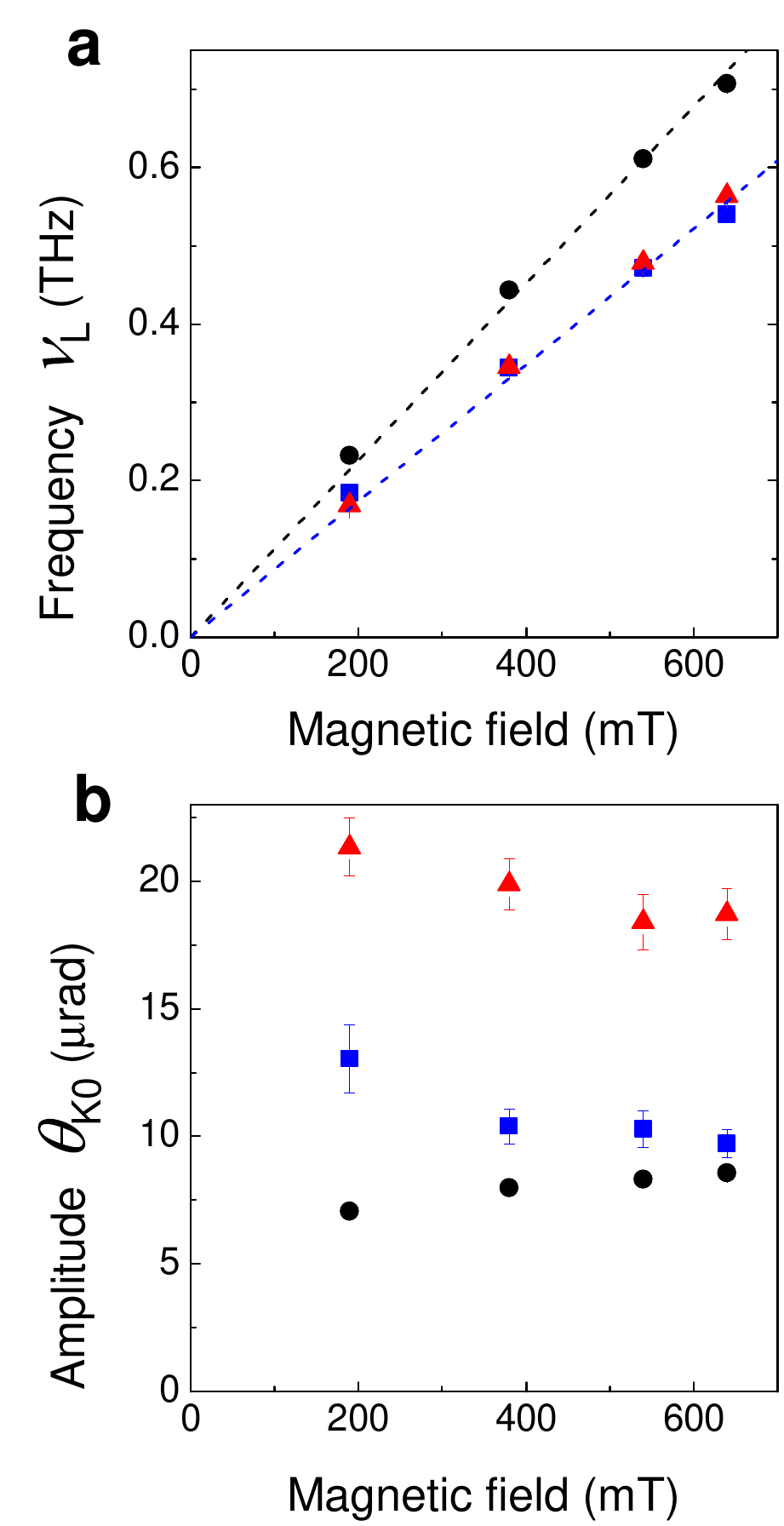}
\caption{\textbf{Magnetic field dependence}  
Magnetic field dependences of the Larmor precession frequency $\nu_L$ (a) and amplitude $\theta_{K0}$ (b) for the bare semiconductor QW (black circles) and hybrid structure using $p$ (blue squares) and $\sigma^+$ (red triangles) polarized excitation. $T=2$~K. All data are taken with a spectrally broad laser without amplitude shaping. For the bare QW structure no bandpass filter has been used in detection.}
\label{Fig:BDep}
\end{figure}

In order to get further insight into the spin initialization of the photoexcited electrons, we analyze the magnetic field dependence of the spin transients. The results are summarized in Fig.~\ref{Fig:BDep}(a) and (b) where the magnetic field dependence of the Larmor precession frequency $\nu_L$ and amplitude are shown, respectively. The data are compared for three different configurations: the bare QW for $\sigma^+$ pump excitation and the plasmonic structure for $\sigma^+$ and $p$-polarized pump excitation. 
In all cases the precession frequency $\nu_L$ grows linearly with $B$. From the slopes of the magnetic field dependences we evaluate the electron $g$-factors. Interestingly, we obtain a larger value $g_e^{\rm QW}=81$ for the bare QW as compared to $g_e^{\rm HP}=62$ in the hybrid plasmonic structure which we attribute to heating effects due to SPP excitation in the metal (for details on the heating of the manganese system see supplementary section 3). We also note that the spin relaxation time $\tau_S \approx 3-5$~ps does not depend on the magnetic field magnitude up to 640~mT or the structure type so that it has to be governed by the interaction with the Mn ions. Indeed, the observed time scale is in good agreement with corresponding previous studies~\cite{Crooker-97}. 
One of the most important results is that the amplitude of the signal $\theta_{K0} \propto S_0$ practically does not depend on magnetic field. Hence, the spin conversion from SPP to exciton is possible even in zero magnetic field. This directly proves that in contrast to the magnetic field induced heavy hole-light hole mixing in QW structures, here it is the optical excitation of light hole excitons that provides the main contribution to the observation of transverse spin orientation via SPP spin fluxes.

In conclusion, we have demonstrated  spin conversion from plasmons to excitons in a planar hybrid semiconductor plasmonic structure, enabling optical orientation of exciton spins by linearly polarized light. The metallic structure is exploited for generation of plasmonic spin fluxes on ultimately short time scales. In our work we have used 30~fs pulses which allows one to perform spin manipulation with THz precision. Even higher spin precession frequencies should be possible for larger magnetic fields. Further, the possibility of spin transfer from the photoexcited carriers to the magnetic ions can be exploited in future. Therefore, the proposed approach for transverse optical spin orientation opens a plenty of new possibilities for triggering spin dynamics in non-conventional geometries, allowing in particular applications in ultrafast spin manipulation by fs optical pulses. 

\vskip 1cm
\noindent{\bf Acknowledgements}

We are grateful to  M.~M.~~Glazov, L.E~Golub, and  A.~M.~~Kalashnikova for useful discussions. We acknowledge the financial support by the Deutsche Forschungsgemeinschaft through the International Collaborative Research Centre 160 (Project C5).  ANP acknowledges the partial financial support from the Russian Foundation for Basic Research Grant No.19-52-12038-NNIO\_a. The research in Poland was partially supported by the Foundation for Polish Science through the IRA Programme co-financed by EU within SG OP (Grant No. MAB/2017/1) and  by the National Science Centre through Grants No. 2017/25/B/ST3/02966 and 2018/30/M/ST3/00276.

\vskip 1cm
\noindent{\bf Additional information}

The authors declare no competing financial interests.

\vskip 1cm
\noindent{\bf Methods}

{\bf Samples} \quad The semiconductor part of the structure was grown by molecular beam epitaxy. The 10-nm-thick diluted-magnetic-semiconductor ${\rm Cd_{0.974}Mn_{0.026}Te}$ QW layer ($E_g = 1.656$~eV at 2~K) was grown after deposition of a 3-$\mu$m-thick ${\rm Cd_{0.73}Mg_{0.27}Te}$ buffer layer ($E_g \approx 2.10$~eV at 2 K) on a (100) GaAs substrate. The QW was covered with a thin 32~nm ${\rm Cd_{0.73}Mg_{0.27}Te}$ cap layer. On top of the semiconductor structure $200 \times 200~\mu$m$^2$ large gold grating areas were patterned, using electron beam lithography and lift-off processing~\cite{TMRLE-2018}. The grating period, slit width, and gold thickness were 250 nm, 50 nm, and 45 nm, respectively. The studied sample contained also areas without gold stripes at the top, which are refereed to as bare QW. The energies of heavy- and light- hole exciton resonances at 1.66~eV and 1.68~eV, respectively, were evaluated from photoluminescence data~\cite{TMRLE-2018} and reflectivity spectra ~\cite{Borovkova-2019} of QWs with the same composition and QW thickness. The sample was loaded into a liquid helium bath cryostat and kept at the temperature of 2~K. Transverse magnetic fields up to 640 mT were applied in the sample plane using a resistive electromagnet (${\bf B} || x$). The sample was oriented with the slits perpendicular to the magnetic field direction ($y$-axis, see Fig.~\ref{Fig:Transients}(a) ).

{\bf Pump-probe studies} \quad The laser pulses were generated by a Ti:Sa oscillator with a central wavelength of 790~nm, a spectral width of about 100~nm and a repetition rate of 80~MHz. The laser beam traveled through a pulse shaper and a compressor which were used to compensate the pulse chirp acquired during propagation through the optical elements before reaching the sample. The compensation was achieved using the multiphoton intrapulse interference phase scan procedure~\cite{MIIPS}. The pulse shaper was equipped with a dual spatial light modulator which allowed us to shape both phase and amplitude of the different spectral components of the laser. In that way we optimized the central wavelength and bandwidth of the laser for resonant excitation of the heavy- and light- hole excitons. Most of the experimental data (unless stated otherwise in the text) were obtained using a shaped laser spectrum with a central wavelength of 756~nm and a full width at half maximum (FWHM) of 40~nm, which corresponds to a photon energy of 1.64~eV with 0.09~eV FWHM. The laser pulse duration at the sample surface was around 30~fs. After passing the pulse shaper and compressor the laser beam was divided by a silica beam splitter into the pump and probe. The temporal delay between the pump and probe pulses was adjusted by a motorized mechanical delay line mounted in the pump beam path. The intensity of the pump beam was modulated using a mechanical chopper at the frequency of 2.1~kHz. 

Polarization optics in both beam paths were used to adjust the intensity and the polarization of the incident laser beams (Glan-Taylor prisms, $\lambda/4$ and $\lambda/2$ retarders). The pump beam polarization before the objective was varied from circular to linear using a $\lambda/4$ plate (see the results in Fig.~4). The pump beam becomes elliptically polarized with degrees of circular and linear polarization $\rho_c=\sin(\alpha)$ and $\rho_l=\cos(\alpha)$, respectively. The main axis of the polarization ellipse is directed along $\alpha$, where $\alpha/2$ corresponds to the angle between the fast axis of the retarder and the polarization axis of the Glan Taylor prism. In some experiments the $\lambda/4$ plate was replaced by a $\lambda/2$ plate when rotation of the polarization plane by the angle $\alpha$ was required for the linearly polarized pump beam (see results of supplementary section 2). The polarization plane of the probe beam was set in the same way using a Glan Taylor prism in conjunction with a $\lambda/2$ plate. The oscillatory signal $\theta_K(t)$ is practically independent on the polarization direction of the probe pulses (see supplementary section 1). Most of the data were measured for the probe polarized along $x$-axis (s-polarized probe beam). 

The laser beams were focused at the sample surface using a single microscope objective ($10\times$ magnification and numerical aperture of $0.26$) into spots with diameters of about 20~$\mu$m. The pulse energy for the pump and probe were set to about 12 and 4~nJ, respectively. The incidence angles for pump and probe were 10$^\circ$ and 7$^\circ$ in the horizontal ($zx$) and vertical ($zy$) planes, respectively as shown in Fig.~\ref{Fig:Transients}(a). The reflected beams were collected by the same objective. The probe beam was guided into the polarization bridge setup equipped with a polarizing beamsplitter (PBS) and balanced photodetectors (PDs) as shown in Fig.~\ref{Fig:Transients}(a). In most of the measurements (except when stated otherwise), an additional interference band-pass filter was introduced into the detection path of the probe beam in front of the PBS. The filter was centered at 740~nm and had a bandwidth of 10~nm. This allowed us to increase the sensitivity by measuring the optical response at the photon energies in the vicinity of the exciton resonance. The PD signal was detected using a lock-in amplifier synchronized at the pump modulation frequency. The polarization rotation angle was calculated as $\theta_K = \frac{\pi}{2}(U_{ac}/4U_{dc})$, where $U_{ac}$ is the amplitude of the ac-signal detected by the lock-in amplifier and $U_{dc}$ is the magnitude of the signal at one of the photodiodes of the balanced receiver.

{\bf Electromagnetic simulations} 
The electromagnetic field at the quantum well has been rigorously calculated numerically using the   scattering matrix technique from ref.~\cite{Whittaker1999}, also known as the~ Fourier modal method~\cite{Weiss09}.
In this approach, the field inside each layer  is decomposed into plane waves, which are coupled due to Bragg diffraction. Next, Maxwell boundary conditions are applied to match the fields in different layers. The Li factorization technique~\cite{Li1996} was used to improve the convergence of the Fourier series.  The  frequency-dependent refractive indices for gold, CdMgTe and GaAs were taken from refs.~\cite{JohnsonChristy,AndreDang,Aspnes}, respectively. We have neglected  the difference between the  refractive indices of  the quantum well and the  CdMgTe spacers around it.


\renewcommand{\thefigure}{S\arabic{figure}}
\setcounter{figure}{0}

\renewcommand{\theequation}{S\arabic{equation}}
\setcounter{equation}{0}

\makeatletter
\renewcommand\@bibitem[1]{\item\if@filesw \immediate\write\@auxout
    {\string\bibcite{#1}{S\the\value{\@listctr}}}\fi\ignorespaces}
\def\@biblabel#1{[S#1]}
\makeatother

\begin{large}
\vspace*{\fill}
\center{\bf Supplementing material}
\vspace*{\fill}
\end{large}

\noindent{\bf 1. Dependence of pump-probe transients for different polarization directions of probe beam}\label{sec: Detection}

\begin{figure}[t!]
\begin{center}
\includegraphics[width=0.5\textwidth]{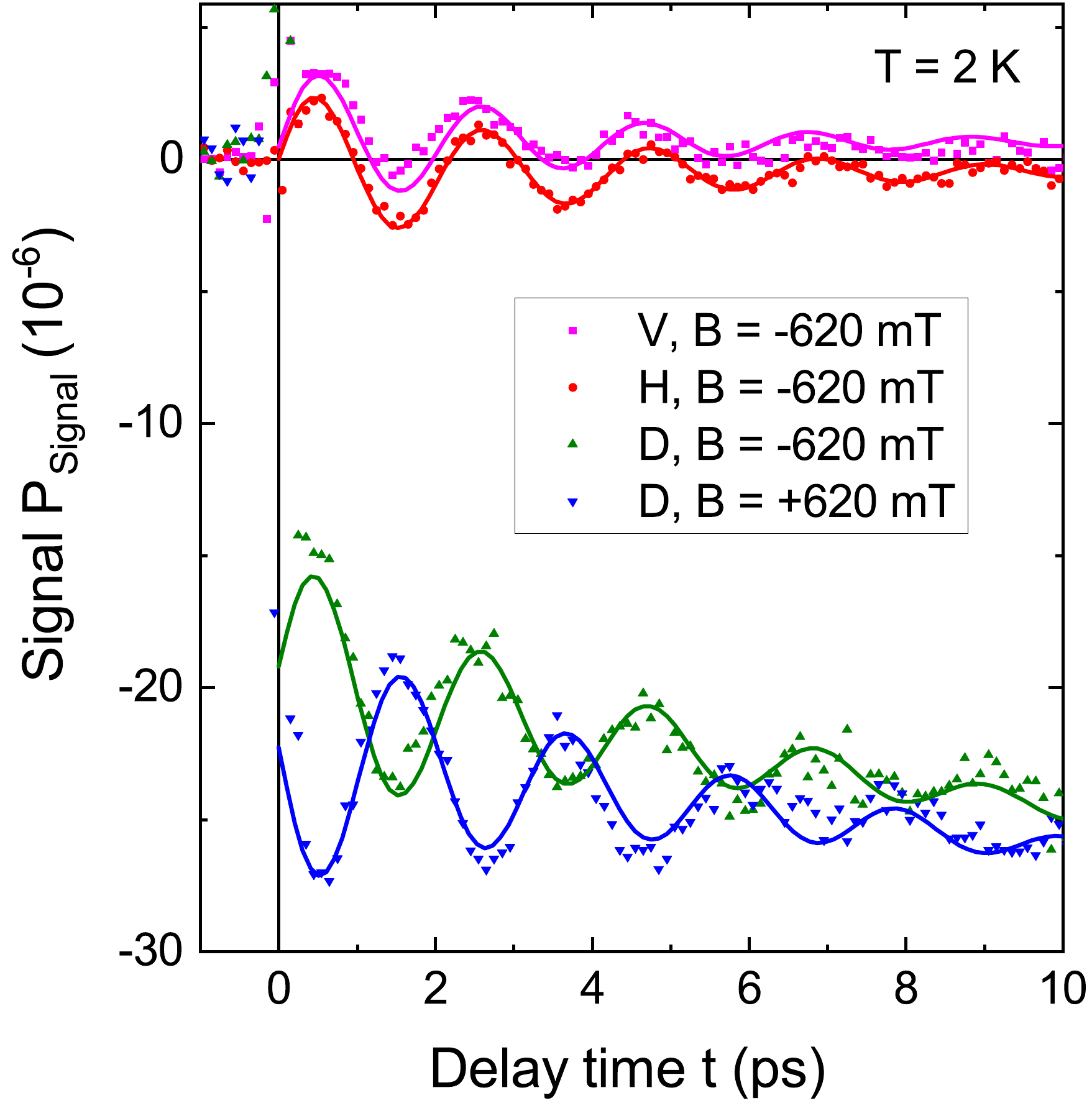}
\end{center}
\caption{Pump-probe transient signals for different directions of polarization plane of probe beam. V and H correspond to p- and s-polarized probe beams, respectively, while the D is the mixture of both. The pump beam is p-polarized. Temperature $T=2$~K.}
\label{fig:Detection}
\end{figure}

Pump-probe transient signal depends on the polarization state of the probe. Figure~\ref{fig:Detection} shows the transients acquired on hybrid plasmonic structure when the probe beam is polarized along $y$ axis (V - vertical), $x$ axis (H - horizontal), 45 deg with respect to $x$ axis (D - diagonal). V and H correspond to p- and s-polarized probe beams, respectively, while the D is the mixture of both. The pump beam is p-polarized perpendicular to grating slits, leading thus to optical orientation of photoexcited electrons via SPPs. The presented signals for $t>0$ can be well approximated by the following expression 
\begin{equation}
P_{\rm Signal}(t) = A+Bt+ \theta_{K0}\cos\left(\frac{2\pi}{T_L}t+\varphi\right)\exp\left(-\frac{t}{\tau_s}\right)
\label{eq:Signal}
\end{equation}
which comprises two contributions. The first one is given by $A+Bt$. It leads to a step-like waveform where the signal jumps from 0 to $A$ at $t=0$ and slowly varies in time which is described by linear dependence with the coefficient $B$. The second contribution is given by the last term on the right side. It leads to an oscillatory signal with the period $T_L$, the phase $\varphi$ and the decay time $\tau_S$. The results of the fits are summarized in  Table~\ref{Tab1}. It follows that the step-like contribution is sensitive to polarization of the probe beam. Its magnitude is nearly zero for p- and s- polarized probe (V,H - polarizations) and appears at strongest for mixed D-polarization. We attribute such behavior to different pump-induced transient response of the reflected probe beam in s- and p- polarizations, e.g. due to SPP resonances which are manifested for H-polarized  probe beam and absent for V-polarized probe in our configuration~\cite{Kreilkamp2016}. For D-polarized probe H- and V- polarizations are detected by different diodes of the balanced detection scheme and therefore the step-like signal reflects the difference $P_{\rm Signal} \propto \Delta R_H - \Delta R_V$, where $\Delta R_H$ and $\Delta R_V$ are the differential reflectivities in H-  and V- polarizations, respectively. This signal may have non-magnetic origin, e.g. changes in the dielectric function of metal or semiconductor layer due to SPP excitation or photoexcited carriers. Therefore we do not discuss this contribution in our work. 

\begin{table}[b!]
\centering
\caption{Main parameters of transient signals measured in different probe polarization configurations (H,V,D)  evaluated from Fig.~\ref{fig:Detection} using Eq.~\ref{eq:Signal}.}
\begin{tabular}{c c c c c c c c}
\hline\hline
Polarization & B(mT) & $A$ ($\mu$rad) & $B$ ($\mu$rad/ps) & $\theta_{K0}$ ($\mu$mrad)  & $T_L$ (ps) & $\varphi$ (rad) & $\tau_S$ (ps) \\
\hline
V & -620 & $0.66$  &  $0$   & $3.0$ & $2.08$ & $-1.63$ & $3.3$  \\
H & -620 & $-0.47$ &  $0$   & $3.2$ & $2.14$ & $-1.42$ & $3.7$ \\
D & -620 & $20$    & $0.46$ & $5.0$ & $2.14$ & $-1.41$ & $3.9$  \\
D & +620 & $22$    & $0.36$ & $5.1$ & $2.11$ & $1.53$  & $4.1$  \\
\hline\hline
\end{tabular}\label{Tab1}
\end{table}

The oscillatory signal is due to Larmor precession of spin-polarized photoexcited electrons around the external magnetic field. This signal is practically independent from the polarization direction of the probe beam. The amplitude $\theta_{K0}$ varies only slightly between 3 and 5~$\mu$rad, the Larmor precession period  $T_L\approx 2.1$~ps, the phase  $\varphi\approx-1.5$ and the electron spin relaxation time $\tau_S \approx 3-4$~ps are independent within the accuracy of the experiment. Note that $\varphi$ changes the sign when magnetic direction is inverted (see the data for D-polarization in Fig.~\ref{fig:Detection} and  H-polarization in Fig.~3c). 

The fact that the oscillatory signal does not depend on polarization direction of probe indicates that it is determined by the polar Kerr effect, i.e. rotation of the polarization probe plane by the angle $\theta_{K}(t)$ due to out-of-plane spin component. We also conclude that the polar Kerr effect makes the only contribution for H- or V-polarized probe into the transient signal $P_{\rm Signal}$ which allows us to probe the spin dynamics of photoexcited electrons along the $z$-axis, $\theta_K(t)\propto S_z$. Here, we discussed the data for hybrid plasmonic structure. For bare QW structure it is well established that polar Kerr rotation plays the dominant role in pump-probe transients~\cite{Crooker-97}.

\noindent{\bf 2. Variation of direction for linear polarization of excitation}\label{sec:LinPol}

\begin{figure}[t!]
\begin{center}
\includegraphics[width=0.9\textwidth]{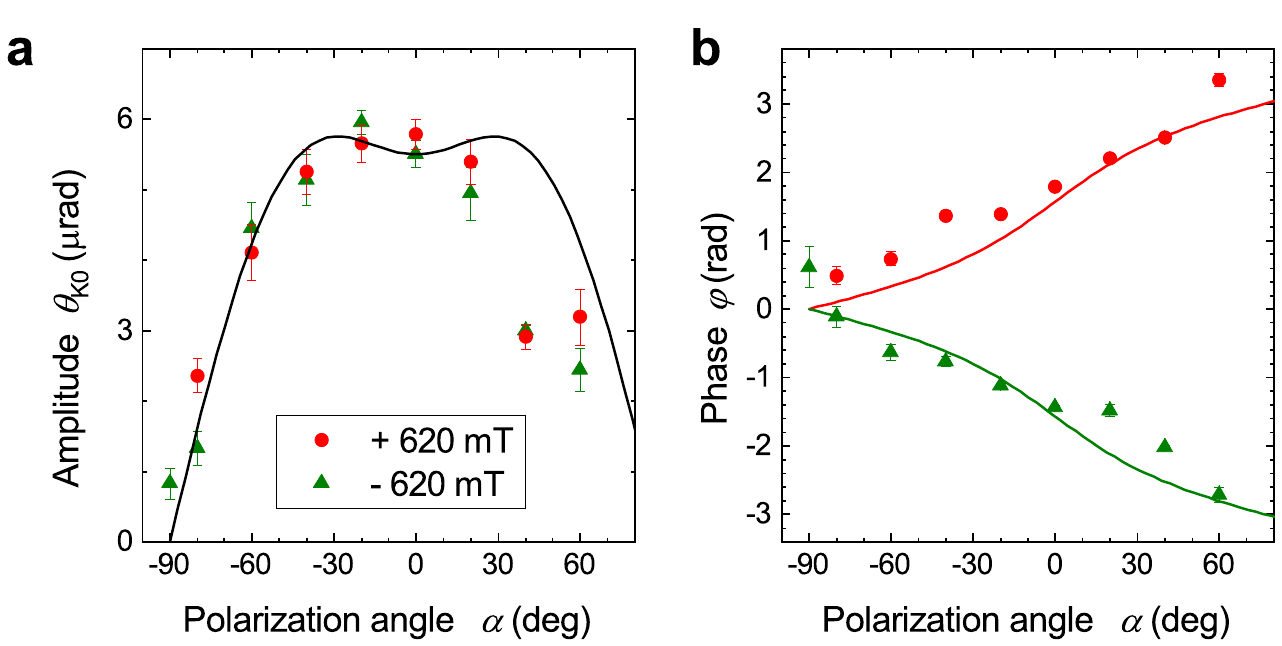}
\end{center}
\caption{Excitation with linearly polarized pump beam. The polarization plane of pump is defined by the angle $\alpha$ which is taken with respect to $x$-axis. (a) and (b) show the dependence of signal amplitude $\theta_{K0}(\alpha)$ and phase $\varphi(\alpha)$, respectively. Dots show the experimental data and solid curves are the results of numerical modeling.}
\label{fig:LinPol}
\end{figure}

Variation of pump polarization changes the relative strength of optical orientation via SPP (near-field effect) or elliptically polarized plane wave transmitted through the grating (far-field effect). This allows one to perform omnidirectional spin control in the $yz$-plane. The variation between circularly and linearly p-polarized pump are discussed in the main text. Here, we consider variation between p- and s- polarized incident pump wave by means of $\lambda/2$-plate. In this case the incident beam is always linearly polarized ($\rho_l=1$ and $\rho_c=0$) and its polarization plane is rotated by angle $\alpha$ with respect to $x$-axis.  $\alpha=0$ corresponds to p-polarized wave which excites SPP with the highest efficiency and $\alpha=\pm90^\circ$ corresponds to s-polarized pump which does not excite SPP. Therefore optical orientation of spin along $y$-axis is present at $\alpha=0$ and absent for $\alpha=\pm90^\circ$. At intermediate angles transmission of pump beam through the grating leads to appearance of elliptically polarized plane wave, which leads to optical orientation of electrons along $z$-axis. Excitation of SPPs is also present even though less efficient. Note, however, that both of the mechanisms vanish for $\alpha=\pm 90^\circ$.  Thus, at intermediate angles $\alpha$ the spin of photoexcited electrons has both $x$ and $y$ components. The ratio between the spin polarization along these axes defines the amplitude $\theta_{K0}$ and phase $\varphi$ of the oscillatory signal 
\begin{equation}
\theta_K(t) \propto S_z(t) = S_{y0}\sin(2\pi\nu_Lt)+S_{z0}\cos(2\pi\nu_Lt)=S_0\cos(2\pi\nu_Lt+\varphi),
\label{eq:Amplitude&Phase}
\end{equation}
with $\theta_{K0}\propto{S_0}=\sqrt{S_{y0}^2+S_{z0}^2}$ and $\varphi=-\arctan(S_{y0}/S_{z0})$. Here, we neglected spin relaxation.

Figure~\ref{fig:LinPol} shows the dependences of $\theta_{K0}(\alpha)$ and $\varphi(\alpha)$. In the vicinity of $\alpha=0$ we obtain large signal amplitude with $\varphi=\pm\pi/2$, i.e. the initial spin polarization is large and directed mainly along $y$-axis. For intermediate angles the contribution of spin polarization along $z$-axis grows, while $S_{y0}$ decreases. In addition the signal amplitude decreases and, as expected the total spin polarization $S_0$ vanishes for $\alpha=\pm90^\circ$. The results of electromagnetic modeling obtained using the same parameters as in Fig.~4c,d  are in good agreement with experimental data.

\noindent{\bf 3. Dependence of spin precession frequency on excitaiton power}\label{sec:Power}

The frequency of Larmor precession $\nu_L$ is sensitive to the temperature of the crystal lattice. The frequency is defined by the magnitude of Zeeman splitting of electron levels 
\begin{equation}
\Delta_e = h\nu_L = x N_0 \alpha \langle S_x^{Mn}(B,T) \rangle,
\label{eq:Zeeman}
\end{equation}
where $x$ is the Mn concentration, $N_0\alpha=0.22$~eV is the exchange constant and steady state spin polarization of the Mn system $\langle S_x^{Mn}(B,T) \rangle$ is defined by the modified  Brillouin function ${\rm B_I}$ for $I=5/2$~\cite{Furdyna,Kossut-book}
\begin{equation}
\langle S_x^{Mn}(B,T) \rangle = S_{eff} {\rm B_{5/2}} \left[\frac{5\mu_Bg_{Mn}B}{2k_BT_{eff}}\right],
\label{eq:Mn-spin}
\end{equation}
where $\mu_B$ is the Bohr magneton, $k_B$ is the Boltzmann constant, $T_c$ is the lattice temperature and $g_{Mn} = 2,01$ is the Mn$^{2+}$ $g$ factor. $S_{eff}$ is the effective spin and $T_{eff} = T_c + T_0$ is the effective temperature. From Eq.~\ref{eq:Mn-spin} it follows that Mn spin polarization and consequently the frequency $\nu_L$ is very sensitive to the lattice temperature. The parameters $x=2.6\%$, $S_{eff}=5/2$ and $T_0=1$~K in the investigated structures were determined in our previous studies~\cite{TMRLE-2018,TMOKE-narrow-2019}. Therefore, it is possible to evaluate the temperature of the crystal lattice $T_c$ in bare QW and hybrid plasmonic parts of the sample, which may vary from the bath temperature $T$ due to optical excitation with intensive laser pulses.

\begin{figure}[t]
\centering
\includegraphics[width=1.15\textwidth]{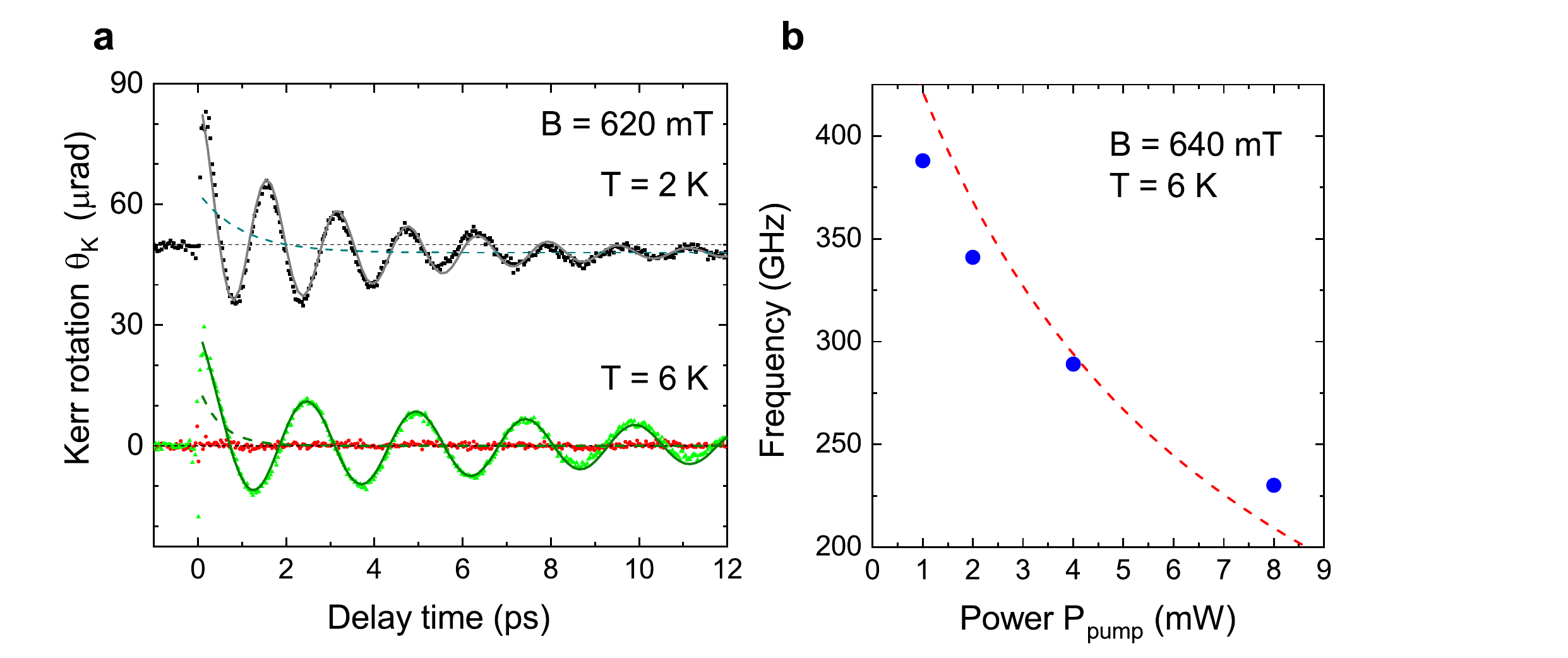}
\caption{Dependence of the spin precession frequency in the bare QW structure. The measurements are taken without using interference bandpass filter in the detection (for details see Methods). (a) Pump-probe transients for $B=620$~mT at temperature $T=2$ (black squares) and 6~K (green triangles). Solid curves are fit with Eq.~\ref{Eq:Signal-bareSC}. Dashed curves show the signal corresponding to the first two terms in Eq.~\ref{Eq:Signal-bareSC}. (b) Dependence of the precession frequency in bare QW structure on the excitation power of pump beam. $T=6$~K, $B=640$~mT. Red dashed line corresponds to expected dependence of $\nu_L$ for additional linear increase of lattice temperature $T_c=T+\gamma P_{\rm pump}$ with $\gamma=1.2$~K/mW.  }
\label{Fig:T-dep}
\end{figure}

The pump-probe transients for bath temperatures of $T=2$ and 6~K are shown in Fig.~\ref{Fig:T-dep}a. Note that excitation with linearly polarized pump (red circles) does not lead to optical orientation in bare SC structure due to very fast decoherence of hole spin. At the same time the signals  for $\sigma^+$-polarized pump are well approximated with the following expression
\begin{equation}
\theta_K = \theta_0 + \theta_{hh}\exp\left(-\frac{t}{\tau_{S,hh}}\right) + \theta_{K0}\cos\left( 2\pi\nu_Lt \right) \exp\left(-\frac{t}{\tau_{S}}\right),
\label{Eq:Signal-bareSC}
\end{equation}
where $\theta_0$ is small offset which changes the sign when the helicity of excitation is changed, i.e. it is related with optical orientation but this part of signal decays significantly longer and does not show fast oscillations. It is likely attributed to optical orientation in GaAs substrate. The second term on the right side is due to fast spin relaxation of the heavy holes which in our case corresponds to about 1 and 0.5~ps at $T=2$ and 6~K, respectively. This contribution is present also in Fig.~3b but it is prominent only in case of bare SC for excitation with circularly polarized light. The oscillatory signal (3rd term on right side in Eq.~\ref{Eq:Signal-bareSC}) is due to the spin precession of photoexcited electrons. As expected the frequency decreases from $\nu_L=0.63$ to 0.40~THz when the bath temperature is increased from 2 to 6~K. We determine the temperature of the crystal lattice $T_c=4.4$ and 7.6~K, respectively.

The fact that $T$ can be smaller than $T_c$ under optical excitation is well known. This is the result of heating due to optical absorption and energy relaxation of photoexcited carriers on phonons. The power dependence of precession frequency is presented in Fig.~\ref{Fig:T-dep}b. While the bath temperature is kept constant ($T=6$~K) the frequency decreases with increase of pump excitation power from 1 to 8~mW leading to  heating of crystal lattice $T_c$ from 8 to 14~K. The linear increase of $T_c$ by 1.2~K/mW with increase of pump power $P_{\rm pump}$ and its influence on the precession frequency $\nu_L$ are shown in Fig.~\ref{Fig:T-dep}b with dashed red line, which is in reasonable agreement with experimental data.

The effects of heating due to optical excitation may differ in bare and hybrid plasmonic structure. The heating in plasmonic structure is expected to be larger due to excitation of plasmons which produces increase of the temperature in the metal layer as well as stronger optical excitation in the semiconductor magnetic QW layer by SPPs. For these reasons we observe the difference in the electron g-factors evaluated from Fig.~5b. The magnitude of electron $g$-factor is given by $g_e=\Delta_e/(\mu_B B)$. We obtain the temperatures of Mn system in bare QW and hybrid plasmonic parts of the sample which correspond to about 3 and 4.5~K, respectively. These data have been measured with the spectrally broad unshaped laser pulses and therefore for excitation with the same power of 1~mW in the pump beam the amount of energy absorbed in  the sample and consequently the heating are weaker if compared to the data from Fig.~\ref{Fig:T-dep} (see also Methods for details). 

\begin{figure}[t]
\centering
\includegraphics[width=\textwidth]{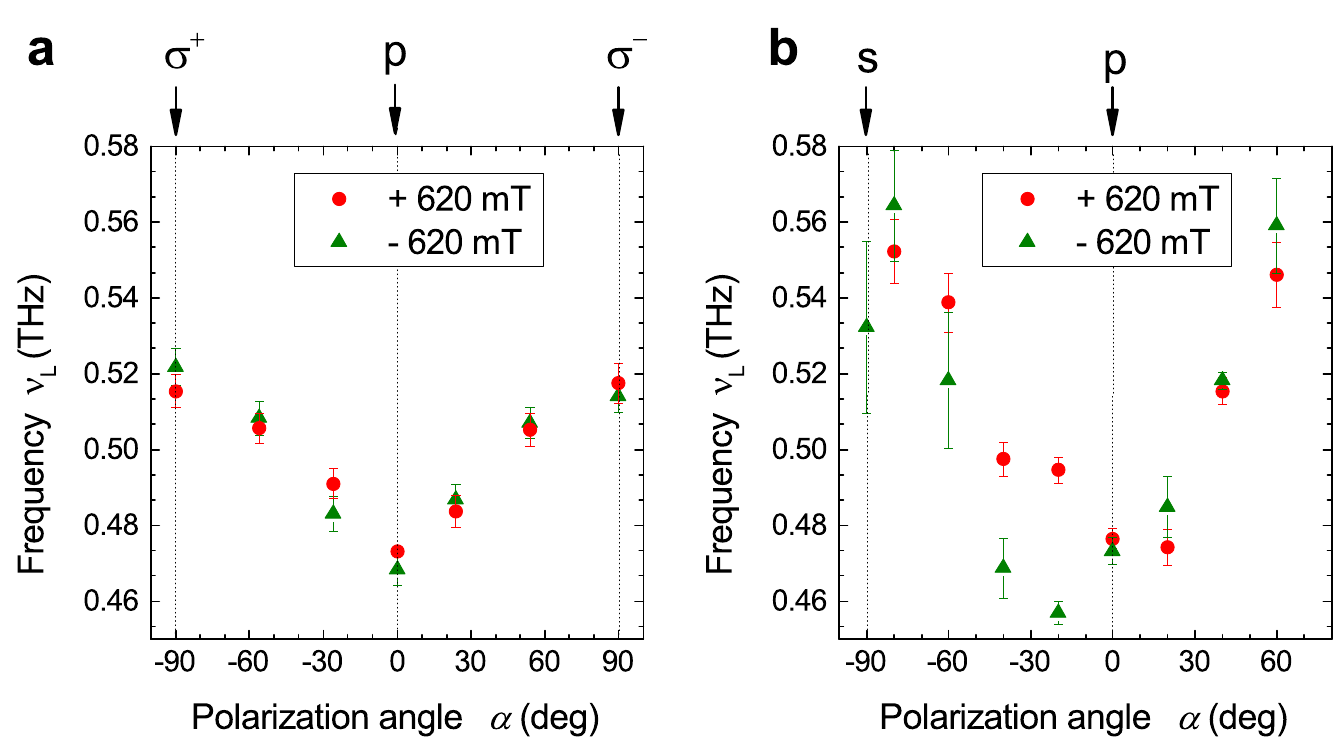}
\caption{The dependence of precession frequency as a function of pump polarization state. The polarization is controlled (a) by $\lambda/4$-plate between circularly $\sigma^\pm$- and linearly p- polarized pump  and (b) by $\lambda/2$ plate between linearly s- and p- polarized pump. The pure polarization states are labeled in the top axis. $B=\pm620$~mT, $T=2$~K.}
\label{Fig:Pola-dep}
\end{figure}

Finally the role of heating due to SPP becomes evident from the dependence of precession frequency on the polarization state of the pump beam, which is shown in Fig.~\ref{Fig:Pola-dep}. The polarization is controlled  by $\lambda/4$-plate between circularly $\sigma^\pm$- and linearly p- polarized pump  (see Fig.~\ref{Fig:Pola-dep}a) or by $\lambda/2$ plate between linearly s- and p- polarized pump (see Fig.~\ref{Fig:Pola-dep}b). The pure polarization states are labeled with arrows in the top axis. The largest $\nu_L$ which corresponds to weak heating is observed for excitation with s-polarized pump when no SPPs are excited. In this case the $\nu_L\approx0.55$~THz, which corresponds to $T_c =5.6$~K, while the bath temperature is kept at $T=2$~K. The lowest $\nu_L\approx 0.47$~THz is evaluated for p-polarized pump beam when SPPs are most efficient. In this case we estimate $T_c=7.2$~K. 

Thus, we conclude that the differences in Larmor precession frequency and consequently the $g$-factors are attributed to stronger heating of Mn system under optical excitation in hybrid structure which is the result of resonant SPP excitation. The heating may appear due to excitation of hot electrons in the Au grating or stronger localization of optical field in the QW structure~\cite{Spitzer-2016}.  This result indicates that SPPs play an important role in heat-assisted processes. 

\noindent{\bf 4. Relative contribution of conventional and SPP mechanisms}\label{sec:Modeling}
\begin{figure}[t]
\centering\includegraphics[width=0.7\textwidth]{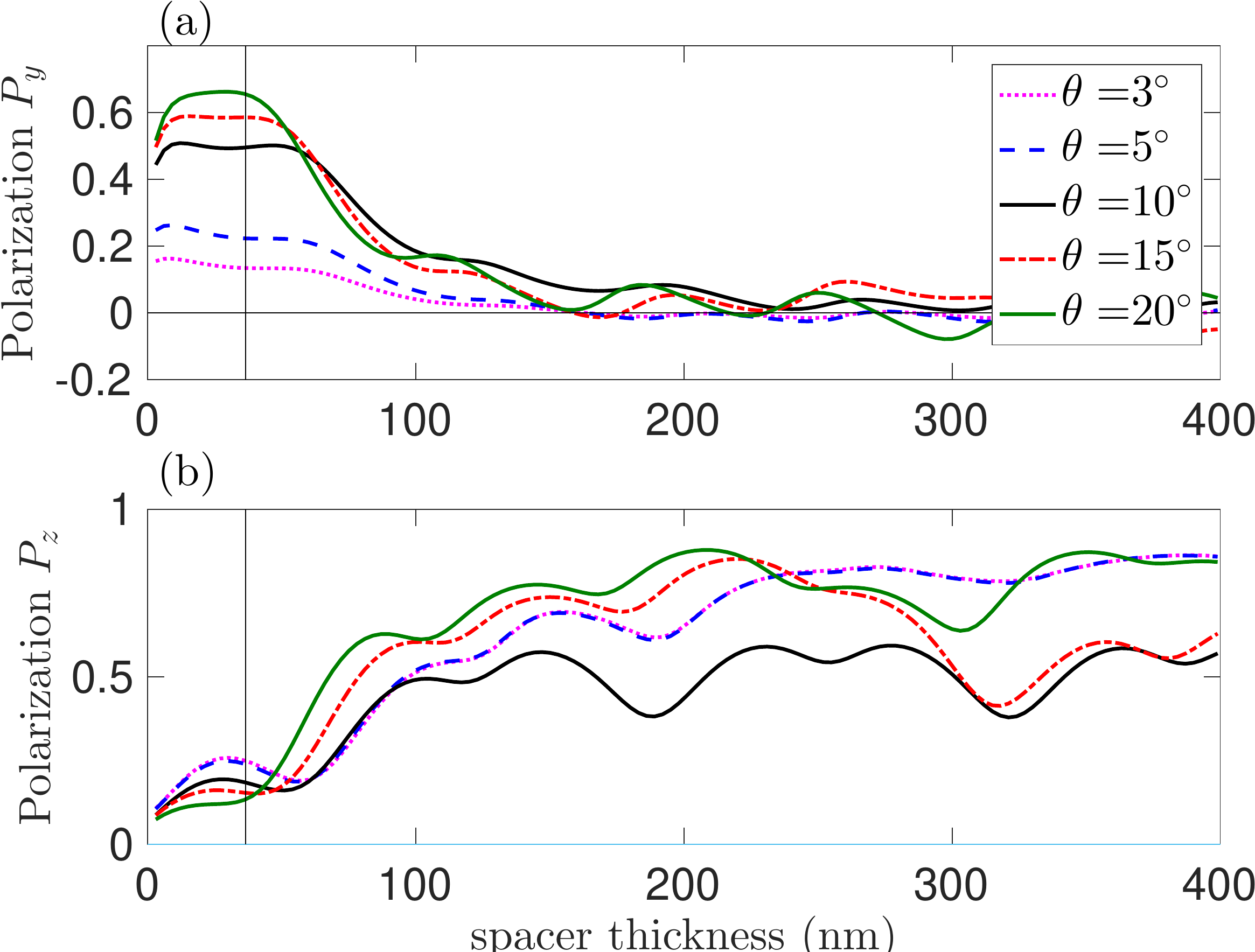}
\caption{Dependence of the average transverse (a) and longitudinal (b) polarizations of electromagnetic field $P_y$ and $P_z$ on the distance from the metallic grating. The structure is excited by a circularly polarized light. Curves have been calculated numerically for different angles of incidence $\theta$ at the same excitation energy $1.7$~eV. Thin vertical line corresponds to the experimental distance of $\approx 37$~nm.}
\label{Fig:th1}
\end{figure}

In order to get more insight into the transformation of light polarization by the  metallic grating, we plot in Fig.~\ref{Fig:th1} the dependence of local electromagnetic field polarization
 \begin{equation}\label{eq:S}
 P_{z}=\frac{2{\rm Im}\: \langle E_xE_y^*\rangle}{\langle |E|^2\rangle},\quad  P_{y}=\frac{2{\rm Im}\: \langle E_zE_x^*\rangle}{\langle |E|^2\rangle}\:,
 \end{equation}
on the thickness of the spacer, that determines the distance from the metallic layer to the quantum well. Here, the angular brackets denote the averaging  over the in-plane coordinate $x$ inside the unit cell of the grating. The structure has been excited by a circularly polarized light at different incidence angles. The idea behind the calculation is that the near electromagnetic field of the plasmons, determining the transverse polarization $P_y$, decays exponentially with the distance from the grating, while the far field effects remain. This is exactly what is seen in Fig.~\ref{Fig:th1}(a): the polarization $P_y$ is suppressed when the distance becomes more than 50~nm. The nonzero oscillating contribution to $P_y$ at the larger distances is related to the far-field effect, namely, the light reflection from the GaAs substrate. Since the transverse polarization is locked to the in-plane wave vector of the plasmons, it vanishes at normal incidence, compare magenta and blue curves in Fig.~\ref{Fig:th1}(a), calculated for $\theta=3^\circ$ and $\theta=5^\circ$. At larger angles the polarization is saturated at the values $P_y\approx 0.6$.

The behavior of the polarization $P_z$ in the out-of-plane direction,  plotted in  Fig.~\ref{Fig:th1}(b), is very different from the transverse one. The calculation demonstrates that the $P_z$ polarization is mostly dictated by the far field effects: it is suppressed in the near field region and gets larger when the distance from the metallic grating increases. The fact that the polarization has also weak oscillations with the distances indicates the presence of the phase shift for $s$- and $p$- components of the far field introduced by the grating. 

\begin{figure}[t]
\centering\includegraphics[width=0.7\textwidth]{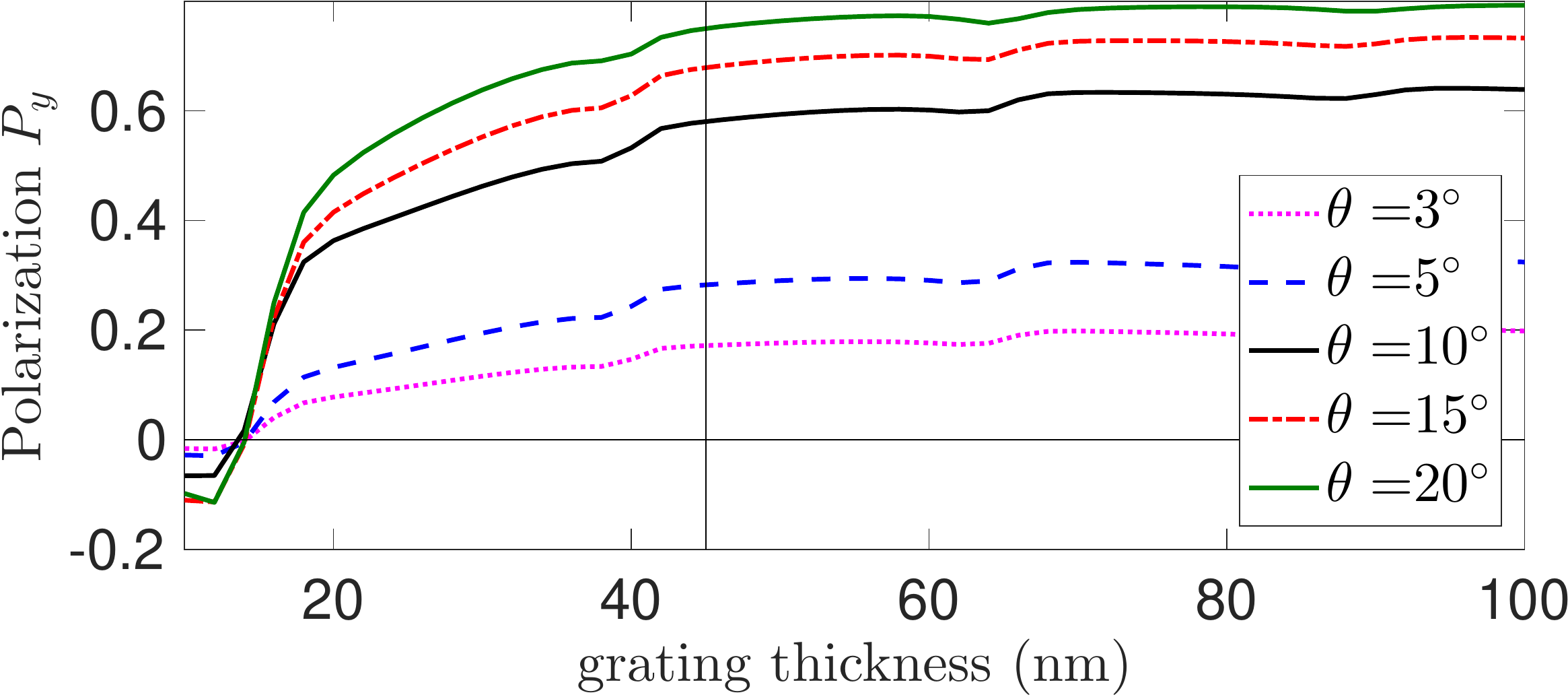}
\caption{Dependence of the average transverse  polarizations of electromagnetic field $P_y$  on the thickness of the  metallic grating. The structure is excited by a $p$-polarized light. Curves have been calculated numerically for different angles of incidence $\theta$ at the same excitation energy $1.7$~eV. Thin vertical line corresponds to the thickness of the grating in the experimental sample  equal to $45$~nm.}
\label{Fig:th2}
\end{figure}

We remind, that while in the near field region the electric field has mostly transverse polarization, $P_y\gg P_z$, the excitation efficiency of the transverse spin is significantly lower, $a_{lh}/a_{hh}=1/\eta\sim 0.1$, because this process involves light-hole states, see also Eq. (2) in the main text.  Thus, it is interesting to examine the possibilities to increase further the transverse polarization.
Besides the shift of  the position of the quantum well and increase of  the angle of incidence, it is also possible  to optimize the grating parameters. Namely,  Fig.~\ref{Fig:th2}  shows that the transverse polarization can be slightly increased for larger  thicknesses of the metallic grating.   
 
\end{document}